\documentclass[final,5p,times, sort&compress, fleqn,twocolumn]{elsarticle}
\journal{Journal of \LaTeX\ Templates}

\usepackage{amsmath}

\begin{document}

\begin{frontmatter}

\title{Bipartite entanglement in the spin-1/2 Ising-Heisenberg planar lattice  \\  
	   constituted of identical trigonal bipyramidal plaquettes\tnoteref{mytitlenote}}
   
\tnotetext[mytitlenote]{The work was financially supported by the Slovak Research and Development Agency under Contract No. APVV-16-0186 and by the Ministry of Education, Science, Research and Sport of the Slovak Republic under Grant No. VEGA 1/0105/20.}

\author{Lucia G\'alisov\'a}
\address{Institute of Manufacturing Management, 
  	     Faculty of Manufacturing Technologies with the seat in Pre\v{s}ov, Technical University of~Ko\v{s}ice, \\
  	     Bayerova 1, 080\,01 Pre\v{s}ov, Slovakia}

\ead{galisova.lucia@gmail.com}

\begin{abstract}
The bipartite entanglement is rigorously examined in the spin-$1/2$ Ising-Heisenberg planar lattice composed of identical inter-connected bipyramidal plaquettes at zero and finite temperatures using the quantity called concurrence. It is shown that the Heisenberg spins of the same plaquette are twice stronger entangled in the two-fold degenerate quantum ground state than in the macroscopically degenerate quantum chiral one. The bipartite entanglement with chiral features completely disappears below or exactly at the critical temperature of the model, while that with no chirality may survive even above the critical temperature of the model. Non-monotonous temperature variations of the concurrence clearly evidence the activation of the entangled Heisenberg states also above classical ground state as well as their re-appearance above the critical temperature of the model.
\end{abstract}


\end{frontmatter}

\section{Introduction}
\label{sec:1}

Entanglement between quantum particles belongs to the most intensively studied topics of today's modern sciences due to its significant role in the development of quantum information processing~\cite{Nie00}. In quantum theory, this quantum-mechanical correlation is investigated mainly in various quantum spin chains~\cite{Arn01,Ost02,Wan02,Zho03,Zha05,Meh14,Wan17,Ana12,Pau13,Str16,Roj17,Kar19,Gal20}, because they are appropriate and simultaneously the simplest candidates for qualitative and quantitative description of the entanglement properties for the purpose of their practical use~\cite{Los98,Bur99}. In addition, quantum spin chains represent a great playground for a comprehensive study of the entanglement at zero as well as finite temperatures. In general, one can say that basic entanglement features in one dimension (1D) are quite well understood by now.

On the other hand, the entanglement between spins in two-dimensional (2D) systems is still not fully explored. Most of theoretical studies are devoted solely to the quantum entanglement in the ground state or at very low temperatures in order to show that this spin correlation may be well tuned by varying the anisotropy parameter~\cite{Gu05,Ros05}, the applied magnetic field~\cite{Syl04,Ros05,Xu10}, as well as by introducing impurities into the system~\cite{Xu10}. However, only a few works deal with thermal entanglement~\cite{Sha07,Ana11,Eki20}. Consequently, the trend of the spin entanglement near continuous (second-order) phase transitions of the 2D quantum spin systems still remains unclear. 

In view of the above facts, the goal of the present Letter is to explore a bipartite ground-state and thermal quantum entanglement in the recently proposed and exactly solved spin-$1/2$ Ising-Heisenberg model on a regular 2D lattice consisting of identical trigonal bipyramidal plaquettes in a zero magnetic field~\cite{Gal19}. The outline is as follows. In Section~\ref{sec:2} we briefly present the magnetic structure of the model and basic steps of its exact analytical treatment. In Section~\ref{sec:3}, the explicit formula for the concurrence determining a bipartite spin entanglement in the Heisenberg triangular clusters is exactly derived and the most interesting numerical results on this quantum-mechanical correlation in the ground state and at finite temperatures are presented in detail. Finally, the most important conclusions are posted in Section~\ref{sec:4}.

\section{Model and its exact treatment}
\label{sec:2}

\begin{figure}[t!]
	\centering
	\includegraphics[width=1.0\columnwidth]{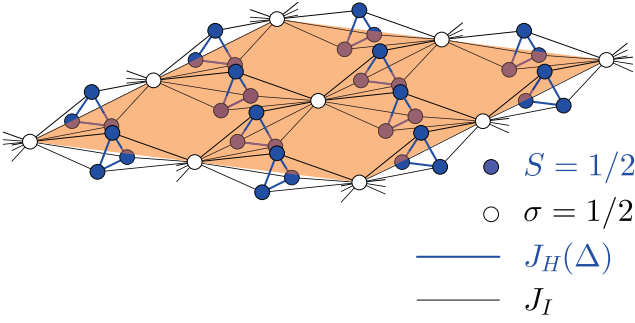}
	\vspace{-5mm}
	\caption{The regular 2D lattice formed by the spin-$1/2$ Ising-Heisenberg trigonal bipyramidal plaquettes. White (blue) circles illustrate lattice sites occupied by the Ising (Heisenberg) spins and thin black (thick blue) lines label the Ising (Heisenberg) exchange interactions. The orange rhombus highlights the lattice plane.}
	\label{fig:1}
\end{figure}

We consider the mixed spin-$1/2$ Ising-Heisenberg model on an infinite regular 2D lattice consisting of $2N$ ($N\to\infty$) identical inter-connected trigonal bipyramidal plaquettes, as schematically depicted in Fig.~\ref{fig:1}.  In this figure, the white circles label lattice sites occupied by the spins $\sigma = 1/2$ which interact with their nearest spin neighbours solely through the Ising-type interaction $J_I$ (thin black line). The blue circles mark lattice sites occupied by the spins $S = 1/2$ which are coupled with each other within the bipyramidal plaquette via the anisotropic XXZ Heisenberg interaction $J_H(\Delta)$ (thick blue line). Under the above assumptions, the total Hamiltonian of the considered 2D mixed-spin model can be written as a sum of $2N$ commuting plaquette Hamiltonians $\hat{\cal H}= \sum_{j=1}^{2N}\hat{\cal H}_j$, where each $\hat{\cal H}_j$ involves all exchange interactions realized within the respective Ising-Heisenberg trigonal bipyramid: 
\begin{eqnarray}
\label{eq:H_j1}
\hat{\cal H}_j \!\!\!\!&=&\!\!\!\!
-J_H\sum_{k=1}^3(\hat{\mathbf S}_{j,k}\cdot\hat{\mathbf S}_{j,k+1})_{\Delta}
- J_{I}\sum_{k=1}^3\hat{S}_{j,k}^{z}(\hat{\sigma}_{j}^z + \hat{\sigma}_{j+1}^z).
\end{eqnarray}
In above, $(\hat{\mathbf S}_{j,k}\cdot\hat{\mathbf S}_{j,k+1})_{\Delta}=\Delta(\hat{S}_{j,k}^x\hat{S}_{j,k+1}^x + \hat{S}_{j,k}^y\hat{S}_{j,k+1}^y)+\hat{S}_{j,k}^z\hat{S}_{j,k+1}^z$, where $\Delta$ is the exchange anisotropy parameter, and $\hat{S}_{j,k}^\alpha$ ($\alpha = x,y,z$), $\hat{\sigma}_j^z$ are spatial components of the spin-$1/2$ operators related to the Heisenberg and Ising spins, respectively. For the sake of simplicity, the periodic boundary conditions $\hat{S}_{j,4}^\alpha \equiv \hat{S}_{j,1}^\alpha$, $\hat{\sigma}_{N+1}^z \equiv \hat{\sigma}_{1}^z$ are assumed for these spins. The letter $J_H$ in the first term of Eq.~(\ref{eq:H_j1}) marks the anisotropic XXZ Heisenberg interaction between the nearest-neighbouring Heisenberg spins, while $J_I$ in the latter term labels the Ising-type interaction between the Heisenberg spins and their nearest Ising neighbours.

\subsection{Diagonalization of the plaquette Hamiltonian}
\label{subsec:2.1}

The specific form of the total Hamiltonian clearly indicates that the exact treatment of the considered spin-$1/2$ Ising-Heisenberg model is conditioned by the diagonalization of the plaquette Hamiltonian~(\ref{eq:H_j1}). The relevant calculation can be performed by introducing two composite spin operators $\hat{\mathbf t}_{j} = \sum_{k=1}^{3}\hat{\mathbf S}_{j,k}$ and $\hat{t}_{j}^{\,z} = \sum_{k=1}^{3}\hat{S}_{j,k}^{z}$, which determine the total spin of the Heisenberg triangle from the $j$th bipyramidal plaquette and its $z$-component, respectively. 
In this representation, the Hamiltonian~(\ref{eq:H_j1}) changes to:
\begin{eqnarray}
\label{eq:H_j2}
\hat{\cal H}_j \!\!\!\!&=&\!\!\!\! \frac{3J_H}{8}(2\Delta+1) - \frac{J_H\Delta}{2}\hat{\mathbf t}_{j}^2 
+ \frac{J_H}{2}(\Delta-1)(\hat{t}_{j}^{\,z})^2 {}
\nonumber\\
\!\!\!\!&&\!\!\!\!{}- J_{I}\hat{t}_{j}^{\,z}(\hat{\sigma}_{j}^z + \hat{\sigma}_{j+1}^z).
\end{eqnarray}
From the physical point of view, the above transcription of the plaquette Hamiltonian corresponds to a rigorous mapping of the spin-$1/2$ Ising-Heisenberg planar lattice onto the equivalent bond-decorated square lattice with nodal sites occupied by the spins $\sigma = 1/2$ and bonds decorated by composite spins which can take two possible values $t=\{1/2, 3/2\}$ referring the total spin of the Heisenberg triangle. It is obvious from Eq.~(\ref{eq:H_j2}) that decorating composite spins of the novel lattice do not directly interact with each other. They are coupled only with their nearest nodal neighbours through the original Ising-type interaction $J_{I}$. On the other hand, the anisotropic XXZ Heisenberg interaction $J_{H}(\Delta)$, which was originally realized within the Heisenberg triangular clusters, now determines the value of an effective single-ion anisotropy acting on composite spins and also shifts the energies corresponding to two different values of these spins.

It is easily to check that the operators $\hat{\mathbf t}_{j}^2$ and $\hat{t}_{j}^{\,z}$ appearing in Eq.~(\ref{eq:H_j2}) satisfy the commutation relations $[\hat{\cal H}_j, \hat{\mathbf t}_{j}^2] = 0$ and $[\hat{\cal H}_j, \hat{t}_{j}^{\,z}] = 0$, which implies that they correspond to conserved quantities with well defined quantum numbers $t_j(t_j+1)$ and $t_{j}^{z} = \{-t_j, -t_j+1,\ldots,t_j\}$ for the given total spin $t_j = \{1/2, 3/2\}$, respectively. Bearing this in mind, the effective Hamiltonian~(\ref{eq:H_j2}) can be immediately put into the fully diagonal form:
\begin{eqnarray}
\label{eq:H_j3}
{\cal H}_j \!\!\!\!&=&\!\!\!\! \frac{3J_H}{8}(2\Delta + 1) - \frac{J_H\Delta}{2}t_{j}(t_{j} + 1) 
+ \frac{J_H}{2}(\Delta-1)(t_{j}^{z})^2 {}
\nonumber\\
\!\!\!\!&&\!\!\!\!{}- J_{I}t_{j}^{z}(\sigma_{j}^z + \sigma_{j+1}^z).
\end{eqnarray}
The diagonalized Hamiltonian~(\ref{eq:H_j3}) can be directly employed for a comprehensive ground-state analysis, as well as a rigorous derivation of the partition function of the spin-$1/2$ Ising-Heisenberg planar model needed for examination of its finite-temperature behaviour.

\subsection{Partition function}
\label{subsec:2.2}

Given the validity of the commutation relation $[\hat{\cal H}_j, \hat{\cal H}_{j^\prime}] = 0$ between different plaquette Hamiltonians~(\ref{eq:H_j1}) [or equivalently effective Hamiltonians~(\ref{eq:H_j2})], the partition function ${\cal Z}$ of the investigated spin-$1/2$ Ising-Heisenberg model can be written in the following partially factorized form:
\begin{equation}
\label{eq:Z1}
{\cal Z} = \sum_{\{\sigma_{j}\}}\prod_{j=1}^{2N}{\rm Tr}_{j}{\rm e}^{-\beta \hat{\cal H}_j} = \sum_{\{\sigma_{j}\}}\prod_{j=1}^{2N}\sum_{t_{j},\,t_{j}^{z}} g_{t_j}{\rm e}^{-\beta {\cal H}_{j}},
\end{equation}
where $\beta = 1/(k_{\rm B}T)$ ($k_{\rm B}$ is Boltzmann's constant and $T$ is the absolute temperature), the summation symbol $\sum_{\{\sigma_{j}\}}$ indicates a summation over all possible spin configurations of the Ising spins $\sigma = 1/2$, the product symbol $\prod_{j=1}^{2N}$ runs over all bipyramidal plaquettes of the original lattice and all bonds of the equivalent bond-decorated square lattice consisting of the composite and standard Ising spins, respectively. The symbol ${\rm Tr}_j$ stands for a trace over all spin degrees of freedom of the Heisenberg spins from $j$th bipyramidal plaquette of the original lattice, while the double summation $\sum_{t_j,\,t_j^z}$ runs over all possible values of the quantum numbers $t_{j}$ and $t_{j}^{z}$ corresponding to composite spins in the equivalent bond-decorated lattice. Finally, $g_{t_j}$ denotes the degeneracy factor, which takes the value $g_{1/2} = 2$ for the quantum number $t_{j} = 1/2$ and $g_{3/2} = 1$ for the quantum number $t_{j} = 3/2$.

It is evident from Eq.~(\ref{eq:Z1}) that individual double summations over the quantum numbers of the composite spins can be performed independently of each others. After performing it, one gains the effective Boltzmann's weight whose explicit form gives the opportunity to use the generalized decoration-iteration mapping transformation~\cite{Fis59,Str10,Roj11}. 
The essence of this algebraic method is to substitute all decorating composite spins and associated interactions by effective Ising-type couplings between remaining nodal spins. 
Particular computational steps including the explicit formula for the effective Boltzmann's weight are listed in our recent paper~\cite{Gal19}.\footnote{Note that there is a typo in Eq.~(4) in our original article; the factor at the beginning of the third line should be $4$.}
The result is the following rigorous mapping equivalence between the partition function ${\cal Z}$ of the spin-$1/2$ Ising-Heisenberg model and the one ${\cal Z}_{I}$ of the uniform spin-$1/2$ Ising square lattice with the temperature-dependent effective nearest-neighbour coupling $J_{ef\!f}$: 
\begin{eqnarray}
\label{eq:Z2}
{\cal Z} (T, J_{I}, J_{H}, \Delta) = A^{2N}{\cal Z}_{I}(T, J_{ef\!f}).
\end{eqnarray}
The mapping parameters $A$ and $J_{ef\!f}$ emerging in Eq.~(\ref{eq:Z2}) are explicitly listed in Ref.~\cite{Gal19}. 

It is worth to note that the mapping relation~(\ref{eq:Z2}) formally closes the rigorous treatment of the investigated spin-$1/2$ Ising-Heisenberg planar model, because the partition function ${\cal Z}_{I}$ of the uniform spin-$1/2$ Ising square lattice is known for years~\cite{Ons44}. 
All important physical quantities gaining an insight into ground-state and finite-temperature features of the model can be calculated from Eq.~(\ref{eq:Z2}), as presented in detail in Ref.~\cite{Gal19}. Some of them, namely the spontaneous magnetization and spatial components of the pair correlation function corresponding to the Heisenberg spins, will be used for a rigorous investigation of the bipartite entanglement between these spins at zero and also finite temperatures.

\section{Bipartite entanglement}
\label{sec:3}

It is evident from the plaquette Hamiltonian~(\ref{eq:H_j1}) that the Heisenberg spins may be entangled only within anisotropic XXZ triangular clusters. The spins of different Heisenberg triangles can never be entangled, because the Ising spins localized at common vertices of the neighbouring trigonal bipyramids represent a barrier for development of any quantum correlation between these spins. 
As a measure of quantum entanglement between two Heisenberg spins of the $j$-th XXZ triangular cluster, we use the quantity called concurrence~\cite{Hil97,Woo98}:
\begin{eqnarray}
\label{eq:C1}
{\cal C}_{k,k+1} = {\rm max}\left\{0, \sqrt{\lambda_{k, k+1}^{(1)}} - \sqrt{\lambda_{k, k+1}^{(2)}} - \sqrt{\lambda_{k, k+1}^{(3)}} - \sqrt{\lambda_{k, k+1}^{(4)}}\,\right\}\!.
\end{eqnarray} 
In above, $\lambda_{k,k+1}^{(i)}$ ($i=1{-}4$) are the eigenvalues of the non-Hermitian matrix $R_{k,k+1} = \rho_{k,k+1}(\sigma^{y}\!\otimes\sigma^{y})\rho_{k,k+1}^{*}(\sigma^{y}\!\otimes\sigma^{y})$ in decreasing order, where 
$\rho_{k,k+1}$ represents the reduced density matrix associated with a pair of the Heisenberg spins at the $k$-th and $(k{+}1)$-st triangle vertices, $\rho^{*}_{k,k+1}$ is the complex conjugate of $\rho_{k,k+1}$, and $\sigma^{y}$ is the Pauli matrix. 
The reduced two-spin density matrix $\rho_{k,k+1}$ can be constructed from the full density matrix $\rho_j = {\rm e}^{-\beta\hat{\cal H}_j}\!/{\rm Tr}_j{\rm e}^{-\beta\hat{\cal H}_j}$ of the $j$-th Heisenberg triangular cluster by tracing over the degrees of freedom of the spin localized at the $(k{+}2)$-nd triangle vertex, i.e., $\rho_{k,k+1} =  {\rm Tr}_{k+2\,}\rho_j$. 
Written in the standard two-spin basis $\{|{+}{+}\rangle_{k,k+1}, |{+}{-}\rangle_{k,k+1}, |{-}{+}\rangle_{k,k+1}, |{-}{-}\rangle_{k,k+1}\}$, it has the form:
\begin{equation}
\label{eq:rho_k}
\rho_{k,k+1} = 
\begin{pmatrix}
u^{+}_{k,k+1}& 0 & 0 & 0\\
0 & v_{k,k+1} & y_{k,k+1} & 0\\
0 & y_{k,k+1} & v_{k,k+1}  & 0\\
0 & 0 & 0& u^{-}_{k,k+1}
\end{pmatrix},
\end{equation}
where individual matrix elements are mixtures of the full density matrix ones. The non-zero elements of the matrix~(\ref{eq:rho_k}) are determined by spatial components of the correlation functions corresponding to the respective spin dimer in the XXZ Heisenberg triangle~\cite{Buk91,Ami04}:
\begin{subequations}
\begin{flalign}
\label{eq:u}
u^{\pm}_{k,k+1}  &=  \frac{1}{4} + \langle\hat{S}_{j,k}^{z}\hat{S}_{j,k+1}^{z}\rangle \pm \langle\hat{S}_{j,k}^{z}\rangle,
\\
\label{eq:v}
v_{k,k+1}  &=  \frac{1}{4} - \langle \hat{S}_{j,k}^{z}\hat{S}_{j,k+1}^{z}\rangle,
\\
\label{eq:y}
y_{k,k+1}  &=  2\langle \hat{S}_{j,k}^{x}\hat{S}_{j,k+1}^{x}\rangle.
\end{flalign}
\end{subequations}
The correlation functions emerging in Eqs.~(\ref{eq:u})--(\ref{eq:y}) are given by Eqs.~(13)--(15) in Ref.~\cite{Gal19}.

Taking into account the above results, the concurrence~(\ref{eq:C1}) quantifying the bipartite entanglement between $k$-th and $(k+1)$-st spins in the $j$-th XXZ Heisenberg triangle can alternatively be expressed as follows:
\begin{eqnarray}
\label{eq:C2}
{\cal C}_{k,k+1} = {\rm max}\left\{0, 4\big|\langle \hat{S}_{j,k}^{x}\hat{S}_{j,k+1}^{x}\rangle\big| - 2\!\sqrt{Q_{k,k+1}}\,\right\}\!,
\end{eqnarray}
where $Q_{k,k+1} = \Big(1/4 + \langle\hat{S}_{j,k}^{z}\hat{S}_{j,k+1}^{z}\rangle\Big)^{2}\! - \langle\hat{S}_{j,k}^{z}\rangle^2$.
It is worth to note that all the transverse as well as longitudinal correlations of the individual spin pairs in the Heisenberg clusters are, in fact, identical, because of the same anisotropic exchange coupling $J_{H}(\Delta)$. This results in the same intensity of the pairwise entanglement of these spins, which is reflected in identical concurrences associated with individual spin pairs: ${\cal C}_{1,2} =  {\cal C}_{2,3} = {\cal C}_{3,1} = {\cal C}$.

\subsection{Bipartite ground-state entanglement}
\label{subsec:3.1}

\begin{figure}[t!]
	\centering
	\vspace{0mm}
	\includegraphics[width=1.0\columnwidth]{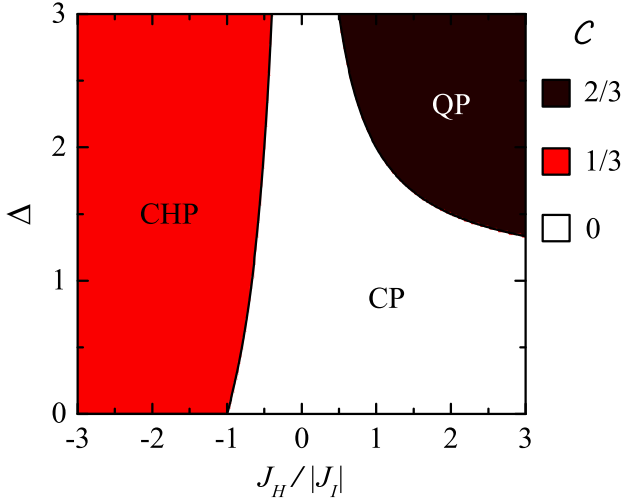}
	\vspace{-6mm}
	\caption{The ground-state phase diagram in the $J_{H}/|J_{I}|-\Delta$ plane supplemented with the zero-temperature density plot of the concurrence ${\cal C}$.}
	\label{fig:2}
\end{figure}

We start with the discussion of the bipartite quantum entanglement in the ground-state phase diagram of the model depicted in the $J_{H}/|J_{I}|-\Delta$ plane in Fig.~\ref{fig:2}. For better clarity, the phase diagram is supplemented by the zero-temperature density plot of the concurrence given by Eq.~(\ref{eq:C2}).
The displayed numerical results are valid for the ferromagnetic ($J_I>0$) as well as antiferromagnetic ($J_I<0$) Ising-type interaction, whereby the transformation $J_I\to-J_I$ causes just a trivial sign change of the Ising spin states, as indicated by respective eigenvectors of three possible long-range ordered ground-state phases, namely:
\begin{itemize}
	\item the classical phase (CP):
\end{itemize}
\begin{equation}
\label{eq:CP}
|{\rm CP}\rangle = \prod_{j=1}^{2N}\!
|{\rm sgn}(J_{I})\rangle_{\sigma_j}\!\otimes |{+}{+}{+}\rangle_j,
\end{equation}
\begin{itemize}
	\item the quantum phase (QP):
\end{itemize}
\begin{equation}
\label{eq:QP}
|{\rm QP}\rangle = \prod_{j=1}^{2N}\!
|{\rm sgn}(J_{I})\rangle_{\sigma_j}\!\otimes\dfrac{1}{\sqrt{3}}\left(
|{+}{+}{-}\rangle_j
\!+\!|{+}{-}{+}\rangle_j
\!+\!|{-}{+}{+}\rangle_j
\right)\!,
\end{equation}
\begin{itemize}
	\item the macroscopically degenerate chiral phase (CHP):
\end{itemize}
\vspace{-3mm}
\begin{eqnarray}
\mbox{\hspace{-2.5cm}}
\label{eq:CHP}
|{\rm CHP}\rangle \!\!\!\!&=&\!\!\!\!\!\prod_{j=1}^{2N}\!
|{\rm sgn}(J_{I})\rangle_{\sigma_j}\!\otimes
 \begin{cases}
\lefteqn{
\frac{1}{\sqrt{3}}\left(
|{+}{+}{-}\rangle_j
\!+\!\omega|{+}{-}{+}\rangle_j
\!+\!\omega^{2}|{-}{+}{+}\rangle_j
\right)
}
\\[3mm]
\lefteqn{
\frac{1}{\sqrt{3}}\left(
|{+}{+}{-}\rangle_j
\!+\!\omega^{2}|{+}{-}{+}\rangle_j
\!+\!\omega|{-}{+}{+}\rangle_j
\right)\,\,
}
\end{cases}
\nonumber\\[-1mm]
&&\hspace{3.55cm} (\omega = {\rm e}^{2\pi{\rm i/3}},\, {\rm i}^2 = -1). 
\end{eqnarray}
In the established notation, the single-site ket vector determines current Ising spin state at $j$-th lattice site ($+$ for $J_{I}>0$ and $-$ for $J_{I}<0$), while three-site ket vector refers to spin arrangement of the neighbouring (also $j$-th) Heisenberg triangular cluster. The sign $+$ ($-$) labels the spin state $1/2$ ($-1/2$) in both kinds of ket vectors. It is worth to note that the spin arrangements described by eigenvectors~(\ref{eq:CP})--(\ref{eq:CHP}) have their energetically equivalent alternatives, which can be obtained by flipping all spin states in individual ket vectors. Owing to this invariance, the CP and QP are two-fold degenerate ground states, while the macroscopic degeneracy of the CHP, caused by two chiral degrees of freedom of each Heisenberg triangle, is $2^{2N+1}$. 

It is obvious from Fig.~\ref{fig:2} and also Eqs.~(\ref{eq:CP})--(\ref{eq:CHP}) that the Heisenberg spins in individual XXZ triangular clusters are quantum-mechanically entangled at zero temperature only in the parameter ranges $J_H/|J_I|>1/(\Delta - 1)$ for $\Delta>1$ and $J_H/|J_I| < -2/(\Delta + 2)$ for any $\Delta>0$, where the two-fold degenerate QP and the macroscopically degenerate CHP are realized as ground states, respectively. However, the  bipartite entanglement observed in these phases is of various strength, as evidenced by different zero-temperature asymptotic values of the concurrence: ${\cal C}=2/3$ for the QP and ${\cal C}=1/3$ for the CHP.
According to these values, any two of the three spins forming the Heisenberg triangles are twice weaker entangled in the CHP than in the QP. The cause is the chirality of the Heisenberg triangular clusters, which weakens the pair coupling between transverse spin components, while the strength of the longitudinal pair coupling as well as the total spontaneous magnetization remain unchanged.

\subsection{Bipartite thermal entanglement}
\label{subsec:3.2}

\begin{figure}[t!]
	\centering
	\vspace{0mm}
	\includegraphics[width=1.0\columnwidth]{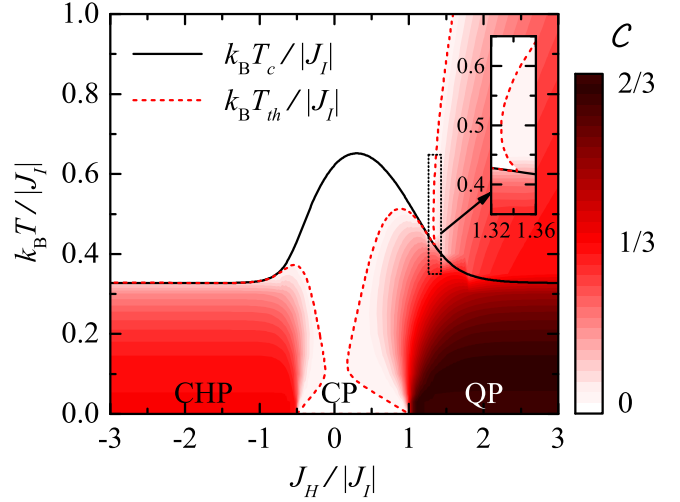}
	\vspace{-6mm}
	\caption{The density plot of the concurrence ${\cal C}$ in the  $J_{H}/|J_{I}|-k_{\rm B}T/|J_{I}|$ plane for the fixed value of the exchange anisotropy $\Delta = 2$. The red dashed line indicates the threshold temperature $k_{\rm B}T_{th}/|J_{I}|$ and the black solid line shows the critical temperature $k_{\rm B}T_c/|J_{I}|$ of the model.}
	\label{fig:3}
\end{figure}

In this part, we take a closer look at the evolution of bipartite quantum entanglement between the Heisenberg spins at finite temperatures. All important information can be read from Fig.~\ref{fig:3} showing the density plot of the concurrence ${\cal C}$ in the $J_H/|J_I|-k_{\rm B}T/|J_I|$ plane for the fixed value of the exchange anisotropy $\Delta = 2$. In this figure, the red dashed and black solid lines indicate the threshold temperature $k_{\rm B}T_{th}/|J_I|$ delimiting the bipartite entanglement and the critical temperature $k_{\rm B}T_{c}/|J_I|$ (second-order phase transition) of the investigated model, respectively. The former temperature was numerically determined from Eq.~(\ref{eq:C2}) by setting ${\cal C}=0$. The latter represents the numerical solution of the exact critical condition $\sinh^2\!\big[J_{ef\!f}/(k_{\rm B}T_c)\big] = 1$ for the uniform spin-$1/2$ Ising square lattice~\cite{Ons44}. Recall that $J_{ef\!f}$ represents the temperature-dependent effective nearest-neighbour interaction given by Eq.~(6) in Ref.~\cite{Gal19}. 

The concurrence and threshold temperature data shown in Fig.~\ref{fig:3} clearly indicate that bipartite entanglements of the Heisenberg spins observed in the zero-temperature parameter regions corresponding to the QP and the CHP persist even at finite temperatures. However, thermal fluctuations generally disrupt quantum spin correlations. As a result, concurrences observed in both ground states usually decrease with increasing temperature until they fall to zero at a certain threshold temperature $k_{\rm B}T_{th}/|J_I|$, which coincides with the critical temperature of the CHP and significantly exceeds the critical temperature of the QP provided the values of the interaction ratio $J_{H}/|J_{I}|$ are taken far enough from the ground-state boundary with the CP (see red dashed lines in Fig.~\ref{fig:3}). This trend can be reversed only around the ground-state boundaries CHP--CP and CP--QP at relatively low negative and positive values of the interaction ratio $-2/(2+\Delta)<J_H/|J_I|<0$ and $0<J_H/|J_I|<1/(\Delta-1)$, respectively, and within a very narrow range of positive values of $J_H/|J_I|$ slightly above the critical temperature of the model, as evidenced by the interesting D- and C-shaped variations of $k_{\rm B}T_{th}/|J_{I}|$ in these particular regions. An unusual temperature-induced generation of bipartite entanglement between the Heisenberg spins just above the classical ground state (CP) as well as its unexpected re-appearing slightly above critical temperature of the model due to further temperature increase can be ascribed to thermal activation of the quantum pair correlations peculiar to the CHP (if $J_H/|J_I|<0$) and/or the QP (if $J_H/|J_I|>0$) in some Heisenberg clusters.

\begin{figure}[th!]
	\centering
	\vspace{0mm}
	\includegraphics[width=1.0\columnwidth]{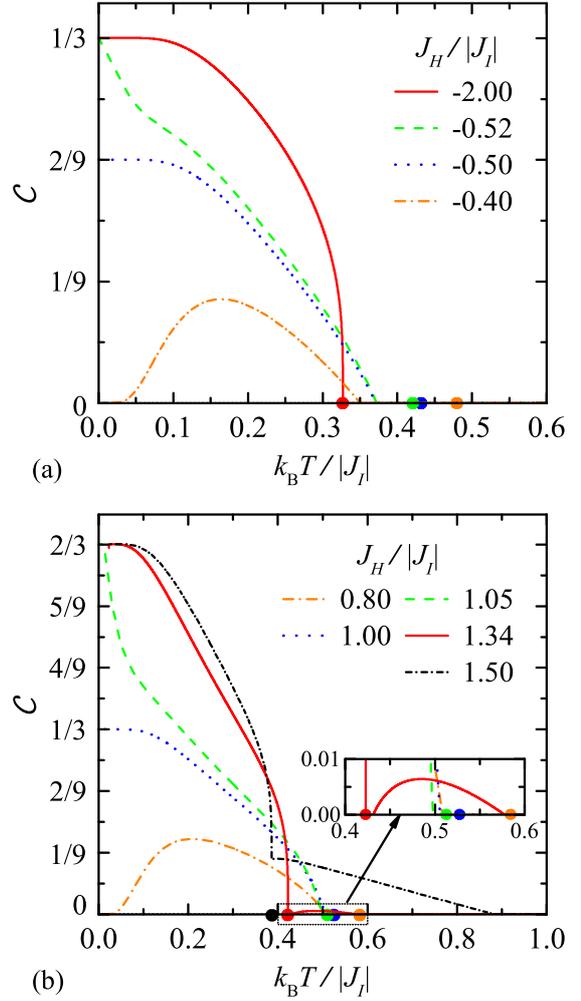}
	\vspace{-6mm}
	\caption{The temperature dependencies of the concurrence ${\cal C}$ for the fixed exchange anisotropy $\Delta = 2$ and several (a)~negative values of the interaction ratio $J_{H}/|J_{I}|$, (b)~positive values of the interaction ratio $J_{H}/|J_{I}|$. The filled circles on temperature axes mark the critical temperatures of the model for given~$J_{H}/|J_{I}|$.}
	\label{fig:4}
\end{figure}

The above findings are demonstrated in more detail in Fig.~\ref{fig:4}, which shows typical temperature variations of the concurrence ${\cal C}$ for a few selected interaction ratios $J_H/|J_I|$ by assuming the same value of the exchange anisotropy $\Delta$ as in Figs.~\ref{fig:2} and~\ref{fig:3}.

Fig.~\ref{fig:4}(a) illustrates the situation when the $J_H/|J_I|$ variation causes the ground-state phase transition between the CHP and the CP. In accordance with previous discussion, the concurrence curve depicted for $J_H/|J_I| = -2.0$ monotonically decreases from the zero-temperature asymptotic value ${\cal C} = 1/3$ with increasing temperature until it rapidly drops to zero at the threshold temperature $k_{\rm B}T_{th}/|J_{I}| \approx 0.327$, which coincides with the current critical temperature of the system. This thermal trend of ${\cal C}$ can be observed whenever the interaction parameters $J_{H}$, $J_{I}$ and $\Delta$ force the system into the macroscopically degenerate CHP, but sufficiently far from the phase boundary with the CP at zero temperature. The concurrence gradually decreases with increasing temperature also for the interaction ratio $J_H/|J_I| = -0.5$ corresponding to the ground-state phase transition CHP--CP. However, it starts from the non-trivial value ${\cal C} = 2/9$ due to coexistence of the entangled chiral Heisenberg spin states with non-entangled ferromagnetic ones until it definitely falls to zero at $k_{\rm B}T_{th}/|J_{I}| \approx 0.372$. The observed threshold temperature is higher than that for $J_H/|J_I| = -2.0$, but still considerably lower than the current critical temperature of the model. The remaining two temperature trends of ${\cal C}$ for $J_H/|J_I| = -0.52$ and $-0.4$ are more pronounced compared to previous ones owing to strong thermally induced excitations to energetically close spin states found in the neighbouring ground states. Namely, in the former case, the concurrence shows a steep decline from the initial value ${\cal C} = 1/3$ at very low temperatures $k_{\rm B}T/|J_{I}| < 0.1$, which can be ascribed to favouring non-entangled ferromagnetic states of the CP in some Heisenberg three-site clusters. In the latter case, the concurrence curve starts from zero, then it shows a vigorous temperature-induced rise starting from the threshold temperature $k_{\rm B}T_{th1}/|J_{I}| \approx 0.028$ and after exceeding a broad maximum at $k_{\rm B}T/|J_{I}| \approx 0.164$ it monotonically decreases with increasing temperature to return to zero at $k_{\rm B}T_{th2}/|J_{I}| \approx 0.351$. The observed non-monotonous trend between $k_{\rm B}T_{th1}/|J_{I}|$ and $k_{\rm B}T_{th2}/|J_{I}|$ clearly demonstrates activation of the bipartite entanglement between the Heisenberg spins present in the macroscopically degenerate CHP above the CP which is stable in the ground state.

More types of temperature variations of the concurrence can be seen in Fig.~\ref{fig:4}(b), which illustrates the scenario where the ground state of the investigated system passes between the CP and the QP with varying $J_H/|J_I|$.
The trends for three lowest values of the interaction ratio $J_H/|J_I|$ are quantitatively similar to those in Fig.~\ref{fig:4}(a). In fact, the concurrence curve depicted for $J_H/|J_I| = 0.8$ shows a single broad peak between two threshold temperatures $k_{\rm B}T_{th1}/|J_{I}| \approx 0.032$ and $k_{\rm B}T_{th2}/|J_{I}| \approx 0.509$, indicating a temperature-induced creation of the entangled QP spin arrangement in the Heisenberg tree-site clusters at the expense of non-entangled CP one present in the ground state. The other value $J_H/|J_I| = 1.0$ causes the phase transition CP--QP at zero temperature in accordance with the ground-state analysis. Hence, the corresponding concurrence curve starts from the mean value of the concurrences identified in the CP and the QP, ${\cal C} = 1/3$, to gradually decrease with increasing temperature until it definitely falls to zero at $k_{\rm B}T_{th}/|J_{I}| \approx 0.506$. The last concurrence curve corresponding to $J_H/|J_I| =1.05$ also shows a gradual decrease to zero, but from the twice higher value ${\cal C} = 2/3$. The observed decrease is followed by a rapid low-temperature decline of ${\cal C}$ due to a massive temperature-induced destruction of the bipartite entanglement characteristic for the QP and favouring the energetically close CP spin arrangement. 
Moreover, Fig.~\ref{fig:4}(b) also illustrates two remarkable concurrence variations for the interaction ratios $J_H/|J_I| = 1.34$ and $1.5$. Both point to the existence of the bipartite entanglement found in the QP not only below but even above the critical temperature of the investigated system. In the former case, the concurrence curve shows a gradual decrease with increasing temperature from the initial value ${\cal C} = 2/3$ until it falls to zero at the first threshold temperature $k_{\rm B}T_{th1}/|J_{I}| \approx 0.423$ coinciding with the critical temperature of the model. 
Surprisingly, it again becomes non-zero at the second threshold temperature $k_{\rm B}T_{th2}/|J_{I}| \approx 0.431$ and definitely drops to zero at the third threshold temperature $k_{\rm B}T_{th3}/|J_{I}| \approx 0.578$ as the temperature further increases [see the detail in Fig.~\ref{fig:4}(b)]. The re-entrance of very weak bipartite entanglement observed slightly above critical temperature of the model can be attributed to the activation of very small population of the entangled QP states. On the other hand, the latter concurrence curve shows a finite cusp at the critical temperature followed by a slight, almost linear decrease to zero at $k_{\rm B}T_{th}/|J_{I}| \approx 0.885$. Thus, the bipartite entanglement of the QP does not completely vanishes exactly at, but far above the critical temperature of the model if the interaction ratio $J_H/|J_I|$ is taken sufficiently far from the ground-state phase transition CP--QP. As demonstrated in Fig.~\ref{fig:3}, the higher the interaction ratio $J_H/|J_I|$ is, the higher the threshold temperature can be found.

\section{Conclusions}
\label{sec:4}

The Letter presents the detailed rigorous study of the bipartite entanglement in the recently proposed and exactly solved regular 2D spin-$1/2$ Ising-Heisenberg lattice composed of identical inter-connected bipyramidal plaquettes in a zero magnetic field~\cite{Gal19}. The quantity concurrence has been used as an indicator for determining a strength of this quantum-mechanical correlation at zero as well as finite temperatures.

It has been shown that the bipartite quantum entanglement between the Heisenberg spins is totally absent at zero temperature only if the spontaneously ordered classical phase with the perfect ferromagnetic arrangement of these spins constitutes the ground state. Otherwise, the Heisenberg spins of the same bipyramidal plaquettes are partially entangled either due to the stability of the two-fold degenerate quantum phase, where these spins are in a symmetric quantum superposition of three possible up-up-down (or down-down-up) states, or the macroscopically degenerate quantum phase characterized by two chiral degrees of freedom of each Heisenberg trimer. The bipartite entanglement observed in the chiral phase is twice weaker than that in the latter phase and completely disappears below or exactly at the critical temperature of the system provided the region far enough from the ground-state boundary with the classical phase. On the other hand, 
the bipartite entanglement of the two-fold degenerate quantum phase may persist even above the critical temperature of the model. In addition, non-monotonous temperature variations of the concurrence clearly prove the activation of the weak bipartite entanglement between the Heisenberg spins also above classical ground state as well its re-appearance slightly above the critical temperature of the model within very narrow range of positive values of the ratio between the Heisenberg and Ising exchange interactions.




\begin{thebibliography}{99}
\bibitem{Nie00} 
M.~A. Nielsen and I.~L. Chuang, \textit{Quantum Computation and Quantum Information}, Cambridge University Press, Cambridge, 2000.
\bibitem{Arn01}
M.~C.~Arnesen, S.~Bose, V.~Vedral, Phys. Rev. Lett. 87 (2001) 017901.
doi:10.1103/PhysRevLett.87.017901
\bibitem{Ost02}
G.~F.~A. Osterloh, L. Amico,  G. Falci, R. Fazio, Nature 416 (2002) 608.
doi: 10.1038/416608a
\bibitem{Wan02}
X. Wang, Phys. Rev. A 66 (2002) 044305.
doi: 10.1103/PhysRevA.66.044305
\bibitem{Zho03}
L.~Zhou, H.~S.~Song, Y.~Q.~Guo, C.~Li, Phys. Rev. A 68 (2003) 024301.
doi:10.1103/PhysRevA.68.024301
\bibitem{Zha05} 
G.~F.~Zhang, S.~S.~Li, Phys. Rev. A 72 (2005) 034302.
doi:10.1103/PhysRevA.72.034302
\bibitem{Meh14}
S.~M.~E.~Mehran, S.~Mahdavifar, R.~Jafari, Phys. Rev. A 89 (2014) 042306.
doi:10.1103/PhysRevA.89.042306
\bibitem{Wan17}
D.~Wang, A.~Huang, F.~Ming, W.~Sun, H.~Lu, Ch.~Liu, L.~Ye, Laser Phys. Lett. 14 (2017) 065203.
doi:10.1088/1612-202X/aa6f85
\bibitem{Ana12}
N.~S.~Ananikian, L.~N.~Ananikyan, L.~A.~Chakhmakhchyan, O.~Rojas, J. Phys.: Condens. Matter 24 (2012) 256001.
doi:10.1088/0953-8984/24/25/256001
\bibitem{Pau13}
H.~G.~Paulinelli, S.~M.~de~Souza, O.~Rojas, J. Phys.: Condens. Matter 25 (2013) 306003.
doi:10.1088/0953-8984/25/30/306003
\bibitem{Str16}
J.~Stre\v{c}ka, R.~C.~Al\'ecio, A.~M.~L. Lyra, O.~Rojas, J. Magn. Magn. Mater.  409 (2016)  124.
doi:10.1016/j.jmmm.2016.02.095
\bibitem{Roj17}
O.~Rojas, M.~Rojas, S.~M.~de~Souza, J.~Torrico, J.~Stre\v{c}ka, M.~L.~Lyra, Physica A 486 (2017) 367.
doi:10.1016/j.physa.2017.05.099
\bibitem{Kar19}
K.~Kar\v{l}ov\'a, J.~Stre\v{c}ka, M.~L.~Lyra, Phys. Rev. E 100 (2019) 042127.
doi:10.1103/PhysRevE.100.042127
\bibitem{Gal20}
L.~G\'alisov\'a, J. Stre\v{c}ka, T. Verkholyak, S. Havadej, accepted for publication in Physica E.
doi:10.1016/j.physe.2020.114089
\bibitem{Los98}
D.~Loss, D.~DiVincenzo, Phys. Rev. A 57 (1998) 120.
doi:10.1103/PhysRevA.57.120
\bibitem{Bur99}
G. Burkard, D.~Loss, D.~P. DiVincenzo, Phys. Rev. B 59 (1999) 2070.
doi:10.1103/PhysRevB.59.2070
\bibitem{Gu05}
S.-J. Gu, G.-S. Tian, H.-Q. Lin, Phys. Rev. A 71 (2005) 052322.
doi: 10.1103/PhysRevA.71.052322
\bibitem{Ros05}
T.~Roscilde, P.~Verrucchi, A.~Fubini, S.~Haas, V.~Tognetti, Phys. Rev. Lett. 94 (2005) 147208. 
doi: 10.1103/PhysRevLett.94.147208
\bibitem{Syl04}
O. F. Sylju{\aa}senu, Phys. Lett. A 322 (2004) 25.
doi: 10.1016/j.physleta.2003.12.018
\bibitem{Xu10}
Q. Xu, S. Kais, M. Naumov, A. Sameh, Phys. Rev. A 81 (2010) 022324.
doi: 10.110PhysRevA.81.022324
\bibitem{Sha07}
S.~E.~Shawish, A.~Ram\v{s}ak, J.~Bon\v{c}a, Phys. Rev. B 75 (2007) 205442.
doi: 10.1103/PhysRevB.75.205442
\bibitem{Ana11}
N.~S.~Ananikian, L.~N.~Ananikyan, L.~A.~Chakhmakhchyan, A.~N.~Kocharian, J. Phys. A: Math. Theor. 44 (2011) 025001.
doi: 10.1088/1751-8113/44/2/025001
\bibitem{Eki20}
C.~Ekiz, J.~Stre\v{c}ka, accepted for publication in Acta Phys. Pol. A. 
\bibitem{Gal19}
L.~G\'alisov\'a, J. Phys.: Condens. Matter. 31 (2019) 465801.
doi: 10.1088/1361-648X/ab37e3
\bibitem{Fis59}
M.~E.~Fisher, Phys. Rev. 113 (1959) 969.
doi:10.1103/PhysRev.113.969
\bibitem{Str10}
J.~Stre\v{c}ka, Phys. Lett. A 374 (2010) 3718.
doi:10.1016/j.physleta.2010.07.030
\bibitem{Roj11}
O.~Rojas, S.~de~Souza, J. Phys. A: Math. Theor. 44 (2011) 245001.
doi:10.1088/1751-8113/44/24/245001
\bibitem{Hil97}
S.~Hill, W.~K.~Wooters, Phys. Rev. Lett. 78 (1997) 5022.
doi:10.1103/PhysRevLett.78.5022
\bibitem{Woo98}
W.~K.~Wooters, Phys. Rev. Lett. 80 (1998) 2245.
doi:10.1103/PhysRevLett.80.2245
\bibitem{Buk91}
D.~J.~Bukman, G.~An, J.~M.~J.~van Leeuwen, Phys. Rev. B 43 (1991) 13352.
doi:10.1103/PhysRevB.43.13352
\bibitem{Ami04}
L.~Amico, A.~Osterloh, F.~Plastina, R.~Fazio, G.~M.~Palma, Phys. Rev. A 69 (2004) 022304. 
doi:10.1103/PhysRevA.69.022304
\bibitem{Ons44}
L.~Onsager, Phys. Rev 65 (1944) 117. 
doi:10.1103/PhysRev.65.117
\end{thebibliography}
\end{document}